\documentclass[fleqn,12pt,twoside]{article}
\usepackage[headings]{espcrc1}
\readRCS
$Id: espcrc1.tex,v 1.2 2004/02/24 11:22:11 spepping Exp $
\usepackage{graphicx}
\usepackage[figuresright]{rotating}

\newcommand \e {\epsilon}

\newcommand \bvec{\left( \begin{array}{c} }
\newcommand \evec{\end{array} \right)}

\newcommand \bea{\begin{eqnarray} }
\newcommand \eea{\end{eqnarray} }

\newcommand {\be} {\begin{equation}}
\newcommand {\ee} {\end{equation}}

\newcommand{\AmS}{{\protect\the\textfont2
  A\kern-.1667em\lower.5ex\hbox{M}\kern-.125emS}}

\title{Hadron-hadron correlations at low and high $p_T$ in heavy-ion collisions}

\author{A.~Majumder\address[DUKE]{Department of physics, Duke University, Box 90305, Durham, NC 27708.}}
\runtitle{Hadron-hadron correlations}
\runauthor{A.~Majumder}

\begin{document}

\voffset -1cm

\maketitle


\begin{abstract}
The modification of two particle correlations within a jet due to its propagation through dense strongly interacting matter is explored. Different properties of the medium may be probed by varying the momentum of the detected hadrons. 
Very high transverse momentum ($p_T$) correlations sample the gluon density of the medium; the minimal modification 
on the same side as the trigger is consistent with the picture and parameters of partonic energy loss.
Lower momentum hadrons, sensitive to the presence of composite structures in the medium may excite collective modes 
such as Cherenkov radiation, resulting in conical patterns in the detected correlations.
\end{abstract}


\section{Introduction}

The goal of ultra-relativistic heavy-ion collisions is the creation and study of excited 
strongly interacting matter, heated past a temperature beyond which confinement may 
no longer be expected~\cite{Jacobs:2004qv}. 
Current experimental results from the Relativistic Heavy-Ion Collider (RHIC) demonstrate 
the formation of matter with bulk collective behaviour that is different from that observed 
in excited hadronic 
matter created at lower energy colliders. Such collectivity, manifested in the radial and elliptic 
flow of the produced matter is also  different from what 
would have been expected from perturbative QCD calculations assuming a quasi-particle 
picture of the quark gluon plasma (QGP).  
The jet correlation analyses reported in these proceedings offers insight on two different 
characteristics of the produced matter.  
Very high momentum partons 
and radiated hard gluons~\cite{Gyulassy:2003mc} selectively sampled 
through the observation of  high transverse momentum hadrons  (referred to as hard-hard correlations) \cite{Adler:2002tq,Adams:2006yt}
tend to sample the partonic substructure of the medium. 
Softer gluons radiated from such hard partons, 
due to their longer wavelengths,  may interact strongly with the prevalent degrees of freedom in the medium. This  
modifies the dispersion relation of the gluons \cite{Koch:2005sx,Majumder:2005sw} 
and may lead to non-jet-like correlations between a high momentum hadron and a softer hadron 
(referred to as hard-soft correlations) \cite{star-cone,Adler:2005ee}. 

\section{Hard-Hard correlations}

\begin{figure}[htbp]
\hspace{1cm}
\resizebox{1.5in}{1.5in}{\includegraphics[1.5in,1.5in][5.5in,5.5in]{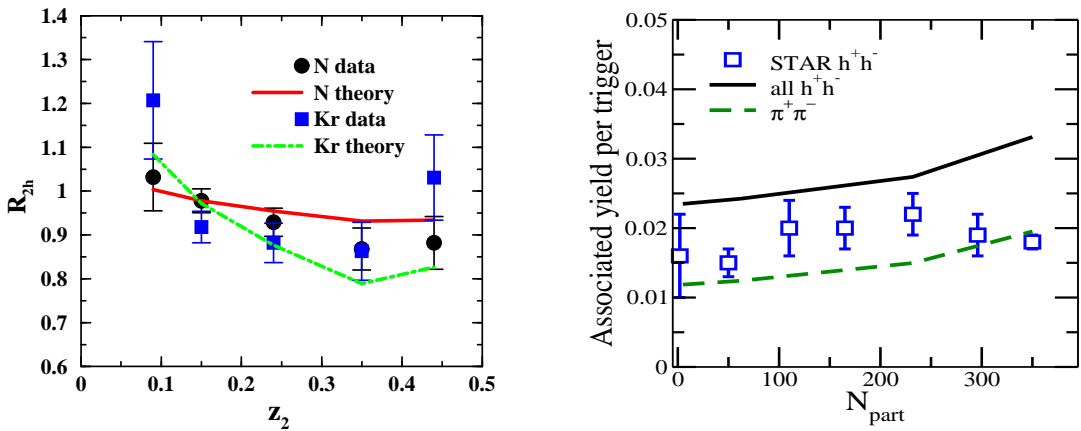}}
\hspace{3cm}
\resizebox{1.5in}{1.5in}{\includegraphics[0.7in,1.2in][4in,5.05in]{associated_yield.eps}}
   \caption{ Results of the medium modification of
    the associated hadron distribution in a cold nuclear medium 
    versus its momentum fraction (left panel) and versus system size in 
a hot medium (right panel) as compared to experimental data (see text for details).}
    \label{fig1}
\end{figure}

We commence with a study of 
the correlations between two hard particles on the same side. 
Such two-hadron correlations have been measured
both in DIS \cite{dinezza04} and high-energy heavy-ion
collisions \cite{Adler:2002tq}. 
Within the energy ranges and angles explored,
it is most 
likely that both these particles have their origin 
in a single jet which is modified by its interaction 
with the medium. Hence, such analysis requires 
the introduction of a dihadron fragmentation function~($D_q^{h_1 h_2}$)
~\cite{Majumder:2004wh}, which accounts for the number of pairs of particles fragmenting from a jet. 
The modified dihadron fragmentation function in a nucleus is calculated in a 
framework similar to that  for the modified single hadron fragmentation
functions \cite{guowang}. As a result, no additional parameters are required.
Due to the existence of sum rules connecting  dihadron fragmentation
functions to single hadron fragmentation functions~($D_q^{h}$)~\cite{Majumder:2004wh}, 
one studies the modification of the conditional distribution for the second rank hadrons,
\begin{eqnarray}
R_{2h}(z_2)\equiv \int d z_1 D_q^{h_1,h_2}(z_1,z_2) \Bigg/ \int d z_1 D_q^{h_1}(z_1),
\label{eq-corr}
\end{eqnarray}
where $z_1$ and $z_2<z_1$ are the momentum fractions of the
triggered (leading) and associated (secondary) hadrons, respectively. 
This associated hadron correlation
is found slightly suppressed in DIS off a nucleus versus
a nucleon target and moderately enhanced or unchanged in central $Au+Au$
collisions relative to that in $p+p$ (in sharp contrast
to the observed strong suppression of single inclusive
spectra in both DIS and central $A+A$ collisions~\cite{hermes1,highpt}).
Shown in the left panel of Fig.~\ref{fig1} is the predicted ratio of the associated
hadron distribution in DIS off a nuclear target ($N$ and $Kr$)
to that off a proton, as compared to the experimental
data from the HERMES collaboration at DESY~\cite{dinezza04}.
The suppression of the associated hadron distribution $R_{2h}(z_2)$
at large $z_2$ due to multiple scattering and induced gluon
bremsstrahlung in a nucleus is quite small compared to the
suppression of the single fragmentation functions~\cite{hermes1,EW1}.
The effect of energy loss seems to be borne, mainly, by the leading
hadron spectrum.

In high-energy heavy-ion (or $p+p$ and $p+A$) collisions, jets are
always produced in back-to-back pairs. Correlations of two
high-$p_T$ hadrons in azimuthal angle generally have two Gaussian
peaks~\cite{Adler:2002tq,Adams:2006yt}. 
The integral of the near-side peak (after background
subtraction) over the azimuthal angle 
can be related to the associated hadron distribution or
the ratio of dihadron to single hadron fragmentation functions [Eq.~(\ref{eq-corr}]. 
The calculation of this distribution requires an integration over the
allowable initial jet energy weighted with the corresponding production
cross sections (obtained via a convolution of the initial structure functions and hard 
partonic cross sections~\cite{Majumder:2004pt,Majumder:2006we}). 
Experimental measurements quote an integrated yield over a range of 
$p_T^{\rm trig}$ and $p_T^{\rm assoc}$, divided by the cross section to produce a trigger hadron, 
integrated over the given range of $p_T^{\rm trig}$. 

Computations of the associated yield, as function of  the number of participants ($N_{part}$) 
are plotted in the right panel of  Fig.~\ref{fig1} along with 
experimental data from Ref.~\cite{Adams:2006yt}.  In this plot, $p_T^{\rm trig}$ ranges from $8-15$ GeV while 
the associated momentum  ranges from $6$ GeV $< p_T^{\rm assoc} < p_T^{\rm trig}$.  Unlike the case in 
DIS (left panel), the data are not normalized by the associated yield in $p+p$ collisions. As a result, the 
normalization is sensitive to the flavour content of the detected hadrons. The experimental results include 
all charged hadrons (with certain decay corrections~\cite{Adams:2006yt}),
 whereas the theoretical predictions include two extreme possibilities: the lower dashed line 
denotes charged pions ($\pi^+,\pi^-$) inclusive of all decays and the upper solid line denotes 
$p,\bar{p},K^+,K^-$ and $\pi^+,\pi^-$ inclusive of all decays. The two cases bracket the experimental 
data, lending support to the framework of partonic energy loss. Decay corrections, which 
essentially involve  removing contributions from unstable particle decays to the detected 
flavour content,  will slightly reduce the plotted associated yields. While no trend may be discerned from 
the experimental measurements, the theoretical predictions show a slight enhancement with centrality due to increased 
trigger bias in more central collisions.

\section{Hard-Soft correlations}

For triggered events in central heavy-ion collisions, the Gaussian peak 
associated with the distribution of high $p_T$ associated particles on 
the away side is almost absent. As the $p_T$ of the associated 
particle is reduced, curious patterns emerge on the away side. A double 
humped structure is seen: Soft hadrons correlated with a quenched jet have a 
distribution that is peaked at a finite angle away from the 
jet \cite{star-cone,Adler:2005ee}, whereas they peak along 
the jet direction in vacuum. The variation of the peak 
with the centrality of the collision indicates that this is not due to 
the destructive interference of the LPM effect. 
Such an emission pattern can be caused by Cherenkov gluon radiation~\cite{Dremin}, 
which occurs only when the permittivity for in-medium gluons becomes larger than unity ($\e > 1$).

\begin{figure}[htbp]
\hspace{1cm}
\resizebox{1.5in}{1.5in}{\includegraphics[1.5in,1.5in][5.5in,5.5in]{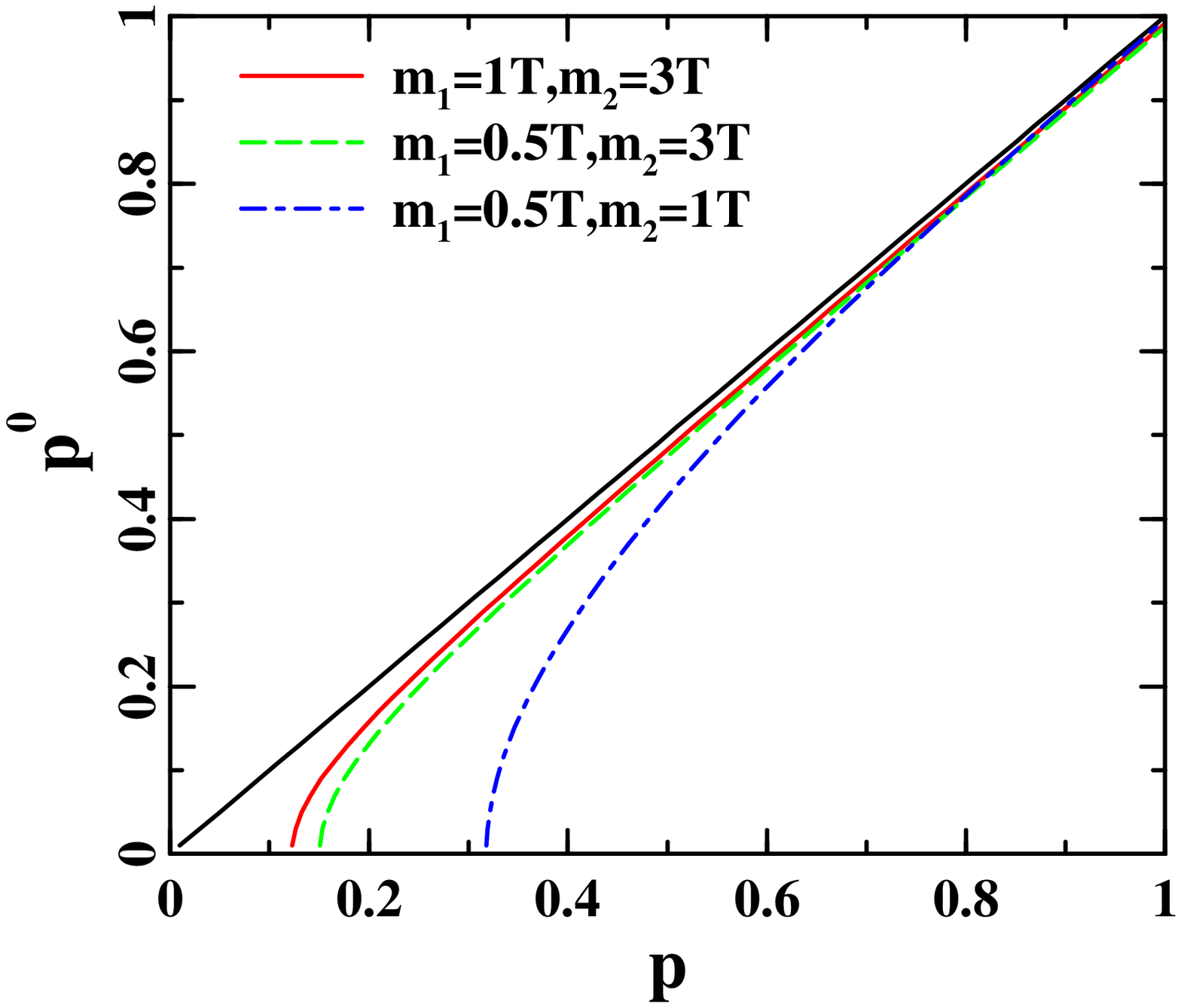}}
\hspace{4cm}
\resizebox{1.5in}{1.5in}{\includegraphics[1.5in,1.5in][5.5in,5.4in]{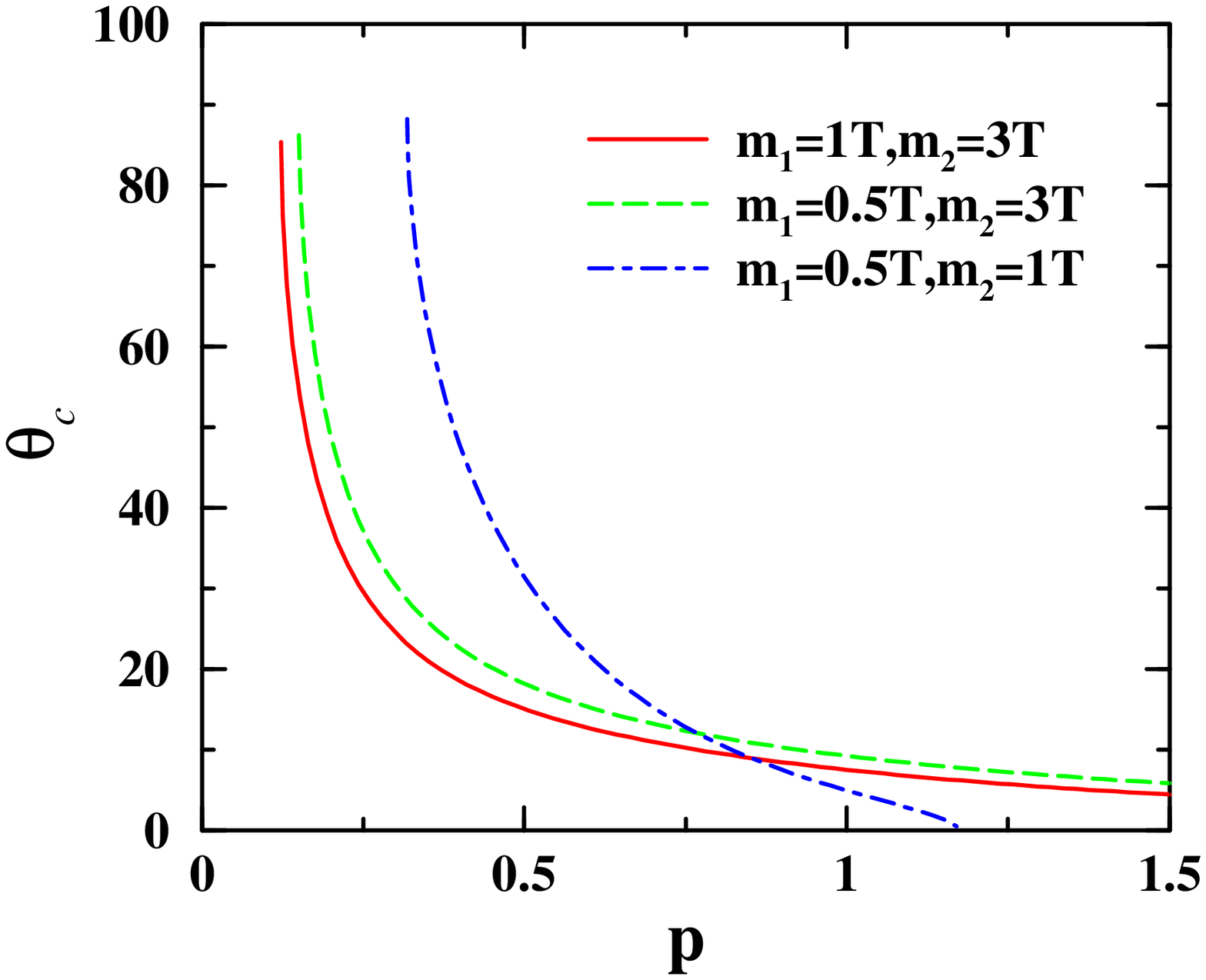}}
    \caption{ The left panel shows the dispersion relation of $\phi$ in a 
thermal medium with transitional coupling to two massive particle. The right panel 
shows the corresponding Cherenkov angles  versus the three momentum of the gluon.}
    \label{fig3}
\end{figure}

The existence of coloured bound states in a deconfined plasma~\cite{Shuryak:2004tx}, 
along with the assumption that these bound states have
excitations which may be induced by the soft gluon radiated from a jet, allows for 
a large index of refraction.
If the energy of the gluon is smaller than that of the first excited state, 
the scattering amplitude is attractive. As a result, the gluon dispersion relation
in this regime becomes space-like ($\e > 1$) and Cherenkov radiation will occur.  
This is simply 
demonstrated in a $\Phi^3$ theory at finite temperature 
with three fields: $\phi$  a massless 
field representing the gluon and two massive fields $\Phi_1$ and $\Phi_2$ 
with masses $m_1$ and $m_2$ in a medium with a temperature $T$ 
(see Ref.\cite{Koch:2005sx} for details). 
The resulting dispersion relations of the $\phi$ field 
for different choices of masses 
of $\Phi_1, \Phi_2$ are shown in Fig.~\ref{fig3} along with the 
corresponding Cherenkov angles of the radiation in the right panels.  
We obtain a space-like dispersion relation at low momentum which approaches 
the light-cone as the momentum of the gluon $(p^0,p)$ is increased. 
The variation of the corresponding angles may be actually detectable in 
current experiments. 
Even though we have studied 
a simple scalar theory, the attraction leading to Cherenkov-like 
bremsstrahlung has its origin in resonant scattering. Thus, the result 
is genuine and only depends on the masses of the bound
states and their excitations. Further experimental and theoretical investigations 
into such correlations may allow for a possible enumeration of the degrees of 
freedom in the produced excited matter. 

Work supported by the U.S. Department of  Energy under grant nos. DE-FG02-05ER41367 and DE-AC03-76SF00098.

\end{document}